\documentclass[
superscriptaddress,
amsmath,amssymb,
aps,
pra,
a4paper,twocolumn,
]{revtex4-2}

\usepackage{bm}

\usepackage[english]{babel}

\usepackage[normalem]{ulem}
\usepackage{bbold}
\usepackage{braket}
\usepackage{xcolor}
\usepackage{csquotes}
\usepackage{appendix}
\usepackage{changes}
\usepackage{array}
\usepackage{hyperref}

\hypersetup{
    colorlinks=true,
    linkcolor=blue,
    filecolor=magenta,      
    urlcolor=cyan,
    anchorcolor = black
    }

\begin{document}

\newcommand{\adri}[1]{\textcolor{teal}{#1}}
\newcommand{\don}[1]{\textcolor{red}{#1}}

\title{Deep Neural Network-assisted improvement of\\ quantum compressed sensing tomography}

\author{Adriano Macarone-Palmieri}\email{amacarone@icfo.net}
\affiliation{ICFO - Institut de Ciències Fotòniques, The Barcelona Institute of Science and Technology, 08860 Castelldefels (Barcelona), Spain}

%
\author{Leonardo Zambrano}
\affiliation{ICFO - Institut de Ciències Fotòniques, The Barcelona Institute of Science and Technology, 08860 Castelldefels (Barcelona), Spain}

\author{Maciej Lewenstein}
\affiliation{ICFO - Institut de Ciències Fotòniques, The Barcelona Institute of Science and Technology, 08860 Castelldefels (Barcelona), Spain}
\affiliation{ICREA, Pg. Lluis Companys 23, 08010 Barcelona, Spain}

\author{Antonio Ac\'in}
\affiliation{ICFO - Institut de Ciències Fotòniques, The Barcelona Institute of Science and Technology, 08860 Castelldefels (Barcelona), Spain}
\affiliation{ICREA, Pg. Lluis Companys 23, 08010 Barcelona, Spain}

\author{Donato Farina}
\affiliation{Physics Department E. Pancini - Università degli Studi di Napoli Federico II,
Complesso Universitario Monte S. Angelo - Via Cintia - I-80126 Napoli, Italy}


\date{\today}

\begin{abstract}
Quantum compressed sensing is the fundamental tool for low-rank density matrix tomographic reconstruction in the informationally incomplete case.
We examine situations where the acquired information is not enough to allow one to obtain a precise compressed sensing reconstruction. In this scenario, we propose a Deep Neural Network-based post-processing to improve the initial reconstruction provided by compressed sensing.
The idea is to treat the estimated state as a noisy input for the network and perform a deep-supervised denoising task.
After the network is applied, a projection onto the space of feasible density matrices is performed to obtain an improved final state estimation. 
 We demonstrate through numerical experiments the improvement obtained by the denoising process and exploit the possibility of looping the inference scheme to obtain further advantages. Finally, we test the resilience of the approach to out-of-distribution data. 
\end{abstract}

\maketitle

\section{introduction}
The measurement and characterization of quantum systems constitute
a long-standing problem in quantum information science \cite{nielsen2010quantum}.
Its complexity originates from the exponential scaling of the Hilbert space dimension $d$ in the size of the system, which in turn implies an exponential scaling for the number of measurements (and copies of the state).
In the most generic case, a complete characterization is required. 
For a $n$-qubit system, $d=2^n$, such full quantum state tomography (QST) necessitates the estimation of $d^2-1=4^n-1$ parameters required to specify the density operator, which is a positive semidefinite (PSD) Hermitian matrix with trace 1.
The limitation of exponential scaling has inspired a wide range of approaches designed to
estimate quantum states with as few measurements
as possible, such as compressed sensing (CS) \cite{CS-Gross2010, CS-Flammia_2012, CS-Riofrio2017, CS-Shikang2019}, adaptive tomography \cite{Adaptive-Granade2017, Adaptive-Husz2012, Adaptive-Lange2023}, matrix product state tomography \cite{MPS-Cramer2010, QST-MPSLanyon2017}, and permutationally
invariant tomography \cite{Permutationally-Toth2010}.
%
These improvements are not free,
as they work under different assumptions on the system state that effectively reduce the number of degrees of freedom involved in the problem.
%

CS approaches are particularly impactful for quantum technology tasks. Indeed, the states in which one is interested for quantum information protocols generally require a high degree of purity, which is exactly the hypothesis needed to successfully run a CS protocol.
Put differently, CS protocols concentrate on cases where the matrix of interest is of low rank, where
the rank is the number of nonzero eigenvalues of the matrix.
In particular, under the assumption that the true state
has (even approximately) rank $r$,
using quantum compressed sensing 
one only needs the knowledge of 
\begin{equation}
\label{cs-scaling}
    O[rd \log^2(d)]
\end{equation}
expectation values of different Pauli operators \cite{CS-Gross2010, CS-Zheng2016}, a remarkable advantage when compared with the general scaling $O(d^2)$, if r is significantly smaller than d.

%

In parallel, recent years have also witnessed a surge of new techniques that are in sync with numerical methods of deep learning (DL) \cite{DNNQST-Carleo2017, DNNQST-Carrasquilla2019, DNNQST-Smith2021, DNNQST-Cha2021, DNNQST-Sotnikov2022, DNNQST-Schmale2022,DNNADAPTIVE-Quek2021}. A handful of solutions have been proposed to handle state reconstruction from the perspective of resource optimization, for example, utilization of incomplete information \cite{DNNQST-Zhu2022, DNNShadow-wei2023, DNNQST-Shahnawaz2021, MEC-Torlai2020, DNNQST-Pan2022} or from the point of view of sample complexity optimization \cite{DNNQST-zhao2023, DNNQST-palmieri2023}.
In particular, the supervised learning approach has recently been shown to be successful in the context of quantum error mitigation \cite{ErrorMit-Bennewitz2022, DNNNERRORMIT-Kim2022} and QST \cite{DNNQST-Palmieri2020}. 
In this regard, we also remark that the supervised deep learning denoiser has been applied for informationally complete (IC) scenarios \cite{DNNQST-palmieri2023} and non-IC, although limited to cat states \cite{DNNQST-Shahnawaz2021}. 
%

Based on these two ingredients, CS and deep neural networks (DNN) tools,
the main objective of our work is to improve CSRs in regimes where the CS protocol is not guaranteed to be successful. This corresponds to the cases where the number $K$ of measured Pauli observables is smaller than the required order \eqref{cs-scaling}.
Interestingly, we can achieve an improvement by concatenating convex optimization techniques and DNN approaches. 
We treat CSR based on trace norm minimization as noisy inputs and apply deep-supervised learning for denoising.
Upon network application, a projection onto the space of feasible density matrices is performed to eventually obtain a physical CSR. Interestingly, the procedure can be iterated, obtaining further improvement.

{
{
}
}

From the machine learning side, the integration of a deep learning component within our protocol offers us the opportunity to study our supervised model to address a broader task than just standard measurement errors for known states; there is also a chance to enhance the reconstruction of target states that the network has not been trained on, which are affected by an unspecified level of depolarizing noise. This brings its application closer to conventional tomography tasks. Remarkably, this improvement does not require the acquisition of new experimental data.
To accomplish this, we employ the \textit{out-of-distribution detection} (OOD) \cite{OOD-Guerin2023,OOD-Nguyen2014} technique. Out-of-distribution is a subfield of ML that focuses on how models perform when they encounter data that are different from the data they were trained on, the latter called the \textit{in-distribution} dataset (ID). In general, this allows us to extend the range of applications of a model and study its resilience to real data. 
This approach has been successfully used with classical models for physical applications, such as anomaly detection for topological phase recognition \cite{ANOMALY-korbinian2020, ANOMALY-Kming2021,ANOMALYVAE-Korbinian2021}, quantum state reconstruction \cite{DNNQST-palmieri2023} or quantum dynamics learning \cite{OUTOFDISTRIB-Caro2023}. We make use of this approach throughout our analysis, showing the potential to reconstruct a greater variety of density matrices outside the ID set, in a different way from the mainstream approaches \cite{DNNQST-Lohani2020}.

The paper is organized as follows.
In Sec.~\ref{sec:prel}
we introduce preliminary notions on quantum compressed sensing and the kind of system and measures that we specify.
In Sec.~\ref{sec:method} we introduce our method, detail how the data are generated, and offer a general introduction to our variational denoising architecture. 
In Sec.~\ref{sec:results} we provide examples of applications for state-of-the-art parallel measurement settings. 
In Sec.~\ref{sec:conc} we draw our conclusions with viable future directions.

\section{preliminaries}
\label{sec:prel}
In this work, we concentrate on $n$-qubit systems. 
The measurements we choose are some of the possible expectation values of Pauli operators (\textit{correlators}, for brevity).
In this regard, we notice that a common decomposition for the density operator ${\rho}$ of an $n$-qubit system is in terms of Pauli operators, 
\begin{eqnarray}
    {\rho}=
\frac{1}{2^n} \sum_{i=0}^{4^n-1} {\rm tr}({O}_i {\rho})\, {O}_i\,,
\label{generalized-bloch}
\end{eqnarray}
where
\begin{equation}    {O}_i:={\sigma}_{i_1}\otimes
{\sigma}_{i_2}\otimes
\dots\otimes
{\sigma}_{i_n}\,,
\label{multiqubitpauli}
\end{equation}
with 
${\sigma}_{i_j} \in \{{\mathbb{1}}_2, {\sigma}_{x}, {\sigma}_{y}, {\sigma}_{z}\}
$. 
The correlator ${\rm tr}({O}_0 \rho)=1$, with ${O}_0:= \otimes_{j=1}^n {\mathbb{1}}_2={\mathbb{1}}_{2^n}$, is fixed because it accounts for the normalization of the state.
This means that, in principle, to unequivocally identify the quantum state, it suffices measure the $4^n-1$ expectation values of the operators \eqref{multiqubitpauli} that appear in the decomposition \eqref{generalized-bloch}.
\subsection{Quantum compressed sensing}
Quantum compressed sensing obtains the reconstruction of a quantum state from the argument of a convex optimization problem \cite{CS-Gross2010}.
Specifically,
\begin{eqnarray}
\begin{array}{rrclcl}
{\rho}_{cs}:= 
\displaystyle {\rm argmin} & \multicolumn{3}{l}{\Vert {\rho}\Vert_{\rm tr}}\\
\textrm{$\rho$ \, s.t.} 
& {\rho}=\rho^\dag\\
& {\rm tr}({\rho})=1\\
& {\rm tr}({O}_i{\rho})={\rm tr}({O}_i{\rho}_{org})\\ &\quad\forall i \in \mathcal{I}.\\
\end{array}
\hspace{2cm}
\label{cs}
\end{eqnarray}
The set $\mathcal{I}$ in \eqref{cs} is intended as
$\mathcal{I}:=\{j_1, j_2, \dots, j_{K}\}$,
which identifies which Pauli operators have been measured, and $\rho_{org}$ is the true \enquote{original} state we want to reconstruct. 
Put differently, given a subset of measurements defined by the set $\mathcal{I}$, compressed sensing provides a state compatible with the information supplied.

From the unicity of the decomposition \eqref{generalized-bloch}, if all Pauli strings are measured, that is, we have $K= 4^n-1$, the fidelity between the true original state and the CS state is 1, i.e., $F(\rho_{org}, \rho_{cs})=1$, where~\cite{nielsen2010quantum}
\begin{equation}
\label{fidelity}
    F(\rho, \sigma)={\rm tr} \sqrt{\sqrt{\sigma}\rho\sqrt{\sigma}}\,.
\end{equation}

In compressed sensing problems one is interested in the non-trivial informationally incomplete regime where $K<4^n-1$.
However, under the assumption that the true state
has, even approximately, rank $r$,
using quantum compressed sensing
to know the expectation values $K_{cs}<4^n-1$ of different Pauli operators
suffices to derive a very good approximation of the unknown state. The needed number of measured correlators
$
K_{cs} 
$
scales in the size of the system as reported in Eq.\,\eqref{cs-scaling} \cite{CS-Gross2010}, providing $F(\rho_{org}, \rho_{cs})\simeq 1$.
This represents a notable advantage compared to the general scaling $O(d^2)$.
On the other hand,
for $K<K_{cs}$ (i.e. outside the validity range of CS), the CSR \eqref{cs} can lead
to an imprecise reconstruction, that is, to a value $F(\rho_{org}, \rho_{cs})<1$. 
This is the regime of interest in the following sections.

\subsection{Denoising with supervised learning}
From the ML point of view, we shall make use and analyze a different application of the \textit{denoising task}, which belongs to the supervised learning class (see, e.g., page 101 in \cite{Goodfellow-et-al-2016} for an introduction to the topic).

We can formalize the {denoising task} as follows. 
We consider a collection of data that we denote as $\boldsymbol{X}$ of elements $x^i\in \mathbb{R}^n$ and a collection $\boldsymbol{Y}$ of target elements $y^i\in \mathbb{R}^n$, where the $x^i$ are \textit{corrupted} (noisy) versions of the targets $y^i$. To perform a denoising task, the algorithm must predict the clean example $y^i$ from the corrupted $x^i$
or, more generally, the conditional probability distribution $P(y^i|x^i)$. This is carried out in the model training phase by minimizing a selected cost function.

\section{The method}
\label{sec:method}

The leading idea behind our approach is to introduce a deep learning step, designed as a standard denoising task, to improve the reconstructions offered by a compressed sensing algorithm. 
We notice that the DNN optimization phase is unconstrained. 
This unavoidably leads to the violation of the PSD requirement of the reconstructed density matrices \cite{REVIEW-Torlai2021}. 
To circumvent this setback, we rely on a convex optimization projection step to find the closest matrix in the feasible set with respect to the Frobenius norm.

\subsection{Data generation}
To realize our data, we employ CSR \eqref{cs} to generate our initial density matrices, inferred from a given subset of known correlators $ \mathcal{I}$. 

First, we notice that, together with the PSD condition, the constraints present in \eqref{cs} identify a feasible convex region $\Omega$, to
which both $\rho_{cs}$ and $\rho_{org}$ belong.
In formulas,  
the feasible set $\Omega$ is defined from
\begin{equation}\label{feas-set}
\rho \in \Omega 
\quad {\rm iff}
\begin{array}{c}
\rho \geq 0\,,\\
{\rm tr}({\rho})=1\,,\\
\qquad
{\rm tr}({O}_i{\rho})={\rm tr}({O}_i{\rho}_{org})
\quad\forall i \in \mathcal{I}\,.
\end{array}
\end{equation}
Note that all states in the feasible region $\Omega$ are CS reconstructions, that is, are optimal for \eqref{cs}, but the solution is not unique, in general.
In other words, solving \eqref{cs} we draw one of these solutions that is not guaranteed to coincide with $\rho_{org}$.
Our work explores the question of whether a nonlinear learning algorithm can pull any CS reconstruction closer to its associated target. 
The complete reconstruction strategy is visualized in Fig.\,\ref{fig:schema}. 

The neural network model applied in our protocol draws inspiration from the supervised approach, where the network function performs a (non-linear) denoising filter \cite{DNNQST-Palmieri2020, DNNQST-Shahnawaz2021, QST-Koutny2022}. 
As such, for each input, we need to associate a target to calculate the cost function.
With this in mind, we build our training dataset $\mathbb{D}$ as follows,
\begin{align}
\label{eq:dataset}
\mathbb{D}=(\boldsymbol{X},\boldsymbol{Y}) :=& \Big\{ \vec{c}_{cs}^{\,j}, \vec{c}_{org}^{\, j} \Big\}_{j=1}^M\,,
\\
{\rm where} \quad 
\vec{c}_{cs/org}^{\,j} :=& \{ {\rm tr}({O}_i \rho_{cs/org}^{\,j})\}_{i=1}^{4^n-1}\,, 
\label{def-ci}
\end{align}
with $M$ the dataset dimension.
The
$\vec{c}_{org}^{\,j} \in \boldsymbol{Y}$ are the target values for the network optimization. Each $\vec{c}_{org}^{\,j}$ contains the $4^n-1$ correlator values (i.e. obtained from all the $O_i$) associated with the density matrix. 
\begin{equation}
\label{haar-states}    
\rho_{org}^{\,j}=\ket{\psi^{j}}\bra{\psi^{j}}~,
\end{equation}
 with $\ket{\psi^{j}}$ a random Haar state \cite{HAAR-Girko1986}.
The $\vec{c}_{cs}^{\,j}\in \boldsymbol{X}$ represents the input of the network. 
A $\vec{c}_{cs}^{\,j}$ is a vector made up of $4^n-1$ correlators, calculated from the density matrix $\rho_{cs}^{\,j}$ obtained from the CSR \eqref{cs}. 
Specifically,
$\rho_{cs}^{\,j}$ is obtained by employing only the measurements defined by the subset $\mathcal{I}$ (see linear constraints in Eq.\,\eqref{cs}), a partial information on the true state $\rho_{org}^{\,j}$. 
Thus, by construction, since both $\rho_{cs}$ and $\rho_{org}$ belong to the feasible set $\Omega$ \eqref{feas-set}, $\vec{c}_{cs}$ and $\vec{c}_{org}$ will differ only for the components that are not listed in the set $\mathcal{I}$.

Last, we anticipate that only for the inference phase we shall consider a more general and realistic quantum state tomography problem,  where our true states are afflicted by some unavoidable noise \cite{Lin2021} and hence might not be pure. We realize it by using depolarized Haar states, 
\begin{equation}
\label{eq:global depo}
    \rho_{org} = (1-p)\ket{\psi}\bra{\psi} + \frac{p}{d}\mathbb{1}\,,
\end{equation}
where $p \in [0,1]$ quantifies the depolarization strength and the degree of mixedness, and the state $\ket{\phi}$ is sampled according to the Haar measure.

\subsection{Model and learning strategy}
In this section, we introduce the architecture we use and the training model.
In our work, the training step is performed with pure Haar states, Eq.~\eqref{haar-states}.

To begin with, we opted to rescale our data from $\mathbb{D}$ to $[0,1]$.
%
This practice is commonly implemented because it may help in gradient optimization during model training. 
%
Apart from this technical detail, the task of the DNN algorithm is to learn the following function,
\begin{align}
\label{eq:network function}
 f_{dnn}:=&[0,1]^{4^n-1}\to [0,1]^{4^n-1}\\
   & \vec{c}_{cs} \to \vec{c}_{den} \simeq \vec{c}_{org}.     \nonumber
\end{align}
With this in hand and a slight abuse of notation, we can frame our problem as a standard denoising problem, where the corruption affects the CSR step \eqref{cs}.

%
\paragraph*{ Architecture.---} The deep learning (DL) architecture we employ in our protocol takes inspiration from a class of models which also combine convolutional layers with an encoder transformer. 
In physics, this class of architectures has recently been applied to extract the characteristics (features) of the classical Brownian motion \cite{REQUENA2023} and for QST \cite{DNNQST-palmieri2023}. Our specific model's equation reads as follows
\begin{equation}
    f_{dnn} = \gamma(l_{cnn})\circ l_{trsf}\circ\gamma(l_{cnn}),
\end{equation}
with $\gamma$ the GELU (Gaussian Error Linear Unit) activation function, which has shown superior performance in models' learning efficiency \cite{GELU-Lee2023},
and $l_{cnn}$, $l_{trsf}$ a convolutional layer and the encoder-transformer layer, respectively.
For model training, we employ a data set with dimension $M= 4\times 10^4$, training/validation partition of $3/1$, batch of dimension 200, and the mean square error cost function.

\subsection{Inference phase}
\label{test}
\begin{figure}
    \centering    \includegraphics[width=0.3\textwidth]{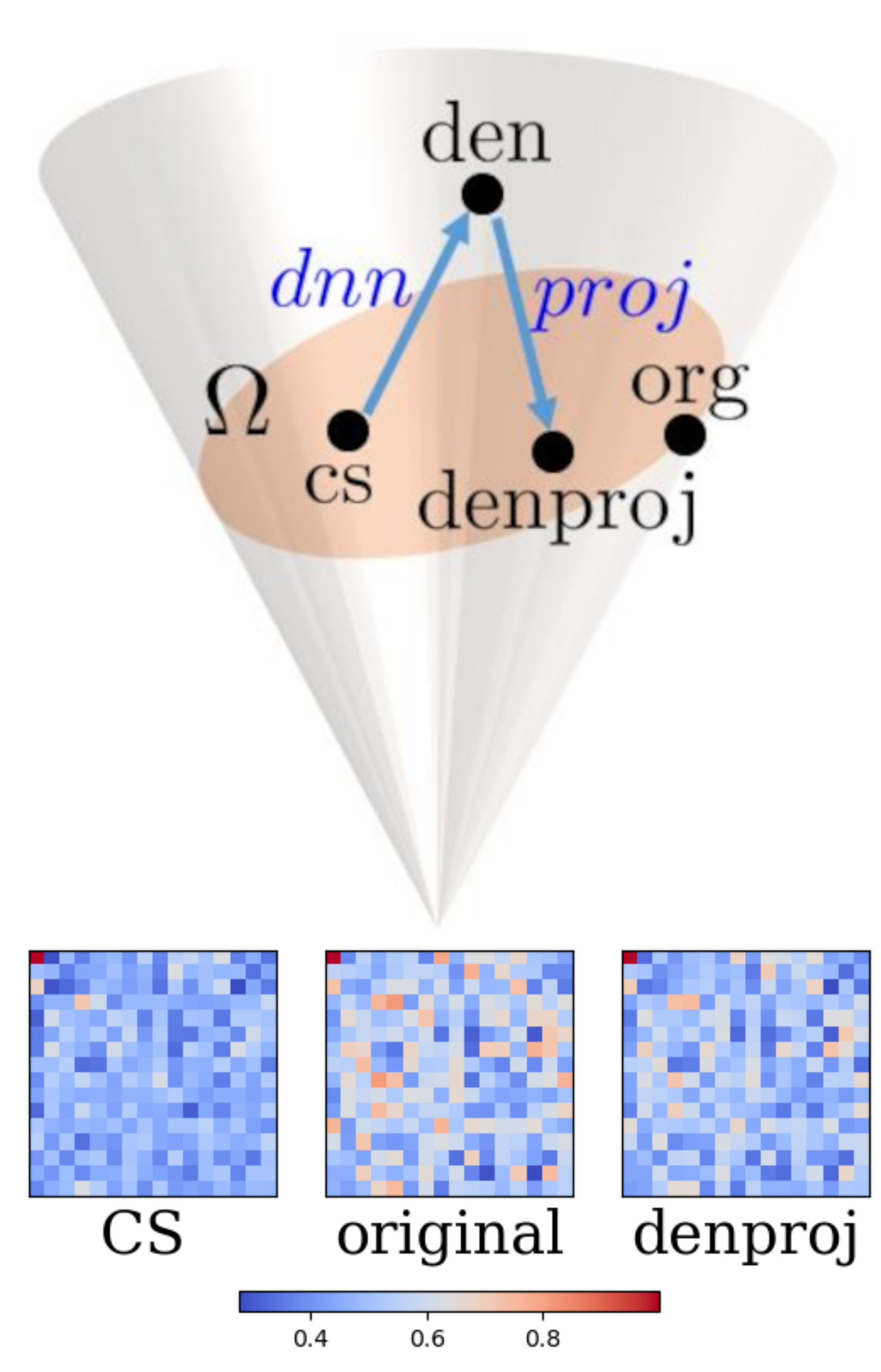}
    \caption{Top: Visual representation of a single step of our inference protocol.
    A CS reconstruction $\rho_{cs}$ is a density matrix within the feasible region $\Omega$ (orange area in the PSD cone).
    The application of $f_{dnn}$ sends the state out of $\Omega$
but provides useful new information on the state.
    A projection is then used to return to the feasible region.
    The aim is to make the resulting state $\rho_{denproj}$ closer to the true state $\rho_{org}$. 
    Bottom: An example of our rescaled input correlators $\vec{c}_{cs}$, the original target $\vec{c}_{org}$, and the reconstruction following projection in the feasible set $\vec{c}_{denproj}$. The images were obtained from the homogeneous set \eqref{homo-setts}.} 
    \label{fig:schema}
\end{figure}

Deep learning-based solutions knowingly suffer from loss of PSD in the outcome reconstructions \cite{REVIEW-Torlai2021}, due to the non-convexity of the optimization problem, steering our reconstructions ${\rho}_{den}$ outside the feasible region $\Omega$. 
Therefore, to complete our protocol, we need to integrate a projection step into the feasible set of valid density matrices. 
For this reason, we add a second step to the inference (test), as Fig.\,\ref{fig:schema} shows.
First, the function $f_{dnn}$ is applied
\begin{equation}
\label{den}
\vec{c}_{den}={f}_{dnn}(\vec{c}_{cs})\,.
\end{equation}

From $\vec{c}_{den}$, after rescaling the coefficients to $[-1,1]$, the (generally, $\not\in \Omega$) state ${\rho}_{den}$ is constructed using Eq~\eqref{multiqubitpauli}.
The closest (physical) state in $\Omega$ is retrieved by performing
the following Euclidean projection,
\begin{equation}
\label{proj}
\qquad \rho_{denproj}=
{\mathcal{P}_\Omega} (\rho_{den})\,,
\end{equation}
where
\begin{eqnarray}
\label{projfull}
\begin{array}{rrclcl}
{\mathcal{P}_\Omega} (\sigma):= 
\displaystyle {\rm argmin} & \multicolumn{3}{l}{\Vert {\rho}-{\sigma}\Vert_{F}}\\
\textrm{$\rho$ \, s.t.} & 
{\rho} \in \Omega\,.
\end{array}
\end{eqnarray}

Upon this step, our test states $\rho_{denproj}$ lie in the feasible region $\Omega$.
With this in hand, 
considering the quantum fidelity \cite{nielsen2010quantum}
\begin{equation}
\label{fidelity}
    F(\rho, \sigma)={\rm tr} \sqrt{\sqrt{\sigma}\rho\sqrt{\sigma}}\,,
\end{equation}
we can now compare
$ F(\rho_{denproj},\rho_{org}) $ and $ F(\rho_{cs},\rho_{org}).
$
To this end, we also introduce the relative gain $\delta f$  for fidelity and $\delta \pi$ for purity, with respect to the sole CS reconstruction as figure of merit, defined as
\begin{eqnarray}
\label{gain}
\delta f&:=&\frac{F(\rho_{denproj},\rho_{org}) - F(\rho_{cs},\rho_{org})}{F(\rho_{cs},\rho_{org})}\\
\delta\pi &:=& \frac{\pi(\rho_{denproj})- \pi(\rho_{cs})}{\pi(\rho_{cs})}
\end{eqnarray}

%

\paragraph*{ Looped inference.---}
Our scheme in Fig.~\ref{fig:schema} suggests that the network optimization direction is slanted toward the feasible set and once projected back into the feasible space, the tangent component shifts our target closer to the targets. Formally, this implies that \eqref{den} to \eqref{proj} can be iterated. 
This prompts us to devise two experiments for the inference phase.
\begin{enumerate}
    \item  We can run an inference step in Eq.~\eqref{den} followed by a projection as in Eq.~\eqref{proj}  several times. In so doing, we have $\vec{ c}_{cs}^{\;(k)}:=\vec c_{denproj}^{\;(k-1)}$, where $\vec c_{denproj}^{\;(k-1)}$ is provided by the previous iteration $k-1$.
    \item  Run a standard inference several times in a row just by applying Eq.~\eqref{den} repeatedly, followed by the final projection in the feasible set to reconstruct our final outcomes.
\end{enumerate}
In this way, we want to unravel to what extent the architecture we use can obtain further improvements inside this protocol, further extending the potential range of application of supervised models.

%

\subsection{Out-of-distribution approach for mixed state reconstruction}
\label{OOT}
In this work, we want to offer an approach that allows us to use a denoising supervised network for the standard QST task, using to our advantage the out-of-distribution detection (OOD) approach. 
According to OOD, the training step is performed with pure Haar states, Eq.~\eqref{haar-states}, while in the test we do not necessarily require the state to be pure, as given by Eq.~\eqref{eq:global depo}.

For a network trained using a collection of states that does not undergo any depolarization, namely $p=0$, we want to unravel to what extent the same model can reconstruct mixed states of lower purity, generated according to Eq.~\eqref{eq:global depo}. 
 
%
In this way, we do not need to retrain the model on the new data. We just reuse a pre-trained model on pure Haar states.

\section{Numerical simulations}
\label{sec:results}

We now show some numerical results that serve to quantify
the effect of the post-processing steps described in Sec.~\ref{test} and Fig.~\ref{fig:schema}.
 
We will consider two cases separately. (i) matrix reconstruction for zero noise, i.e. considering test states as in Eq.\,\eqref{haar-states}, (ii) addition of physical noise through a global depolarizing channel, for different strength values $p$, i.e. considering test states as in Eq.\,\eqref{eq:global depo}.

In our numerical experiments, we consider the states of $n=4$ qubits and the 3 \textit{homogeneous} measurement settings 
\begin{equation}
\label{homo-setts}
    (X^{\otimes 4}, Y^{\otimes 4},Z^{\otimes 4})\,.
\end{equation}
From the marginals of these measurement settings, we can produce a total of $\vert\mathcal{I}\vert=45$  correlators, which are used inside the CS protocol. 
In all our analysis, we consider ideal projectors, remembering that the upper bound of the number of correlators needed is $K_{cs}\simeq 64$ for four qubit pure states \cite{CS-Zheng2016,REVIEWCS-Rani2018}.

%

\subsection{Pure state reconstruction}

To begin with, we analyze our protocol for the reconstruction of pure target states, $\rho_{org}$ as in \eqref{haar-states}. This makes what we call the ID test, when the architecture is trained and tested on the same type of data.

\begin{figure}[h!]
    \centering
    \includegraphics[width = 0.85\linewidth]{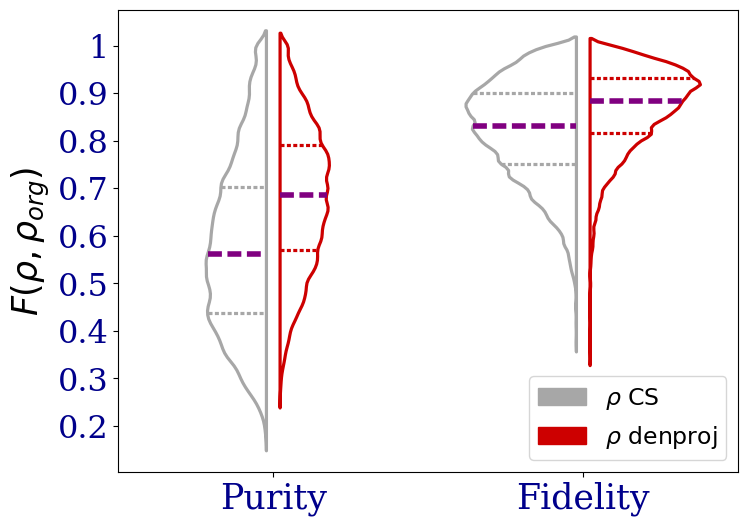}
    \caption{Comparison of the distributions of the quantum fidelity values $F$ and purity values $\pi$ for the input states $\rho_{cs}$ (gray lines) and the projected states upon network reconstruction $\rho_{denproj}$ (red lines). We sample a total of 6000 Haar test states. 
    Throughout the protocol, we used 45 correlators ${O}_i$, derived from the homogeneous set in Eq.~\eqref{homo-setts}. Horizontal dashed purple lines show the average value, dotted gray ones the distributions quartile.
    With our protocol we can obtain an average advantage of $\sim 13\%$ for the purity and $4.7 \%$ for the fidelity, respectively.
    For another example realized using a set of random correlators, see Appendix~\ref{app:pseudorandom}. 
    }
    \label{fig:zero depo homogeneous}
\end{figure}


This first study analyzes the distributions of the values $F$ and $\pi$ for the 6000 Haar test states for the ID test, comparing the input states $\rho_{cs}$ (gray lines) and the reconstructed states $\rho_{denproj}$ (red lines). The findings, illustrated in Fig.~\ref{fig:zero depo homogeneous} and supported by Eq.~\eqref{proj}, demonstrate that the network reconstructions significantly enhance the quality, increasing the average fidelity by approximately $4.7\%$ and purity by approximately $13\%$. This confirms the efficacy of using a supervised denoising strategy to improve CS algorithm reconstructions. Further analysis with random correlators, detailed in Appendix~\ref{app:pseudorandom}, show similar improvements, suggesting that the method's effectiveness is not limited by specific choices of correlators.

\subsection{Mixed state reconstruction}
{

}

%
In this section, we expand the previous denoising, i.e. ID test, to a more general state-tomography reconstruction problem, exploiting the OOD approach; simply by reusing a model previously trained on the dataset given in Eq.~\eqref{eq:dataset} we show that we can still improve the reconstruction performance also for depolarized states of different strengths $p$. We wish to remark that, in so doing, the model is completely agnostic to the amount of noise inside the mixed states. 

\begin{figure}[h!]
    \centering
    \includegraphics[width = 0.9\linewidth]{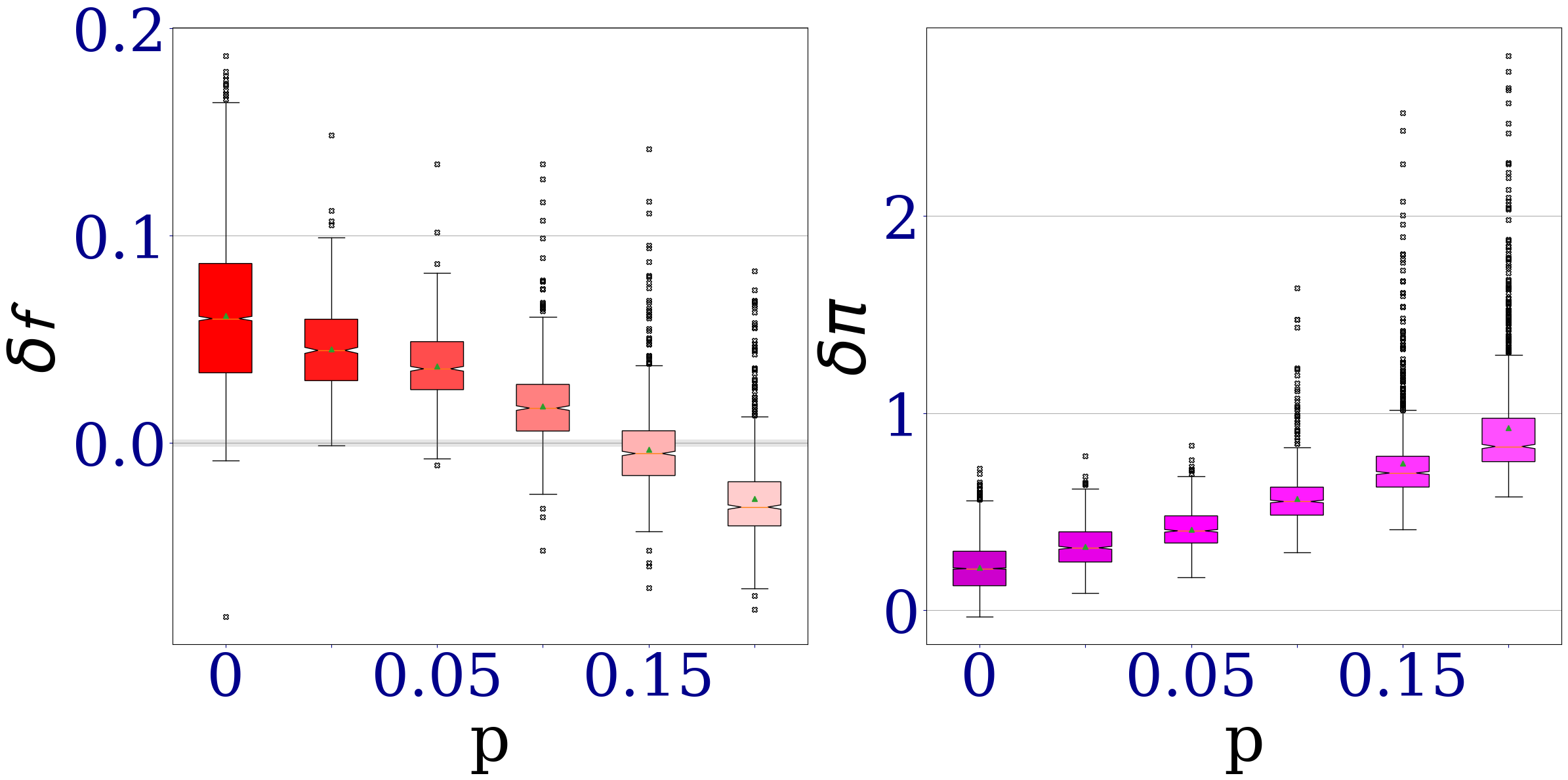}
    \caption{ OOD analysis for 5 different test datasets carried out using the model $f_{dnn}$ trained on non-depolarized states $p=0$. We repeatedly used the same model in inference on six different test datasets of 1000 states each, generated using depolarization strength values of $p= 0,0.025,0.5,0.1,0.15,0.2$, with $p=0$ for the ID dataset. On the left, the quantum fidelity gain $\delta f$ and on the right, the purity gain $\delta\pi$ for each test dataset. Regarding the gain in fidelity, we obtain an advantage for the first four values of $p$, with a $\sim 5.6\%$ for non-depolarized states (the ID testset); then the OOD reconstruction fails for $p = 0.15,0.2$ obtaining a $-0.4\%,-2.8\%$, respectively. The skewness of the distributions increases together with $p$, toward higher values of $\delta f$. In this figure, we used the homogeneous measurement set \eqref{homo-setts}.}
    \label{fig:mixed states}
\end{figure}

In Fig.~\ref{fig:mixed states} we can see the ID test (zero depolarization) together with 5 OOD tests for the state that undergoes a depolarization strength of $p = [ 0.025, 0.05, 0.1,0.15,0.2]$, respectively. 
As the left panel displays, we obtain $\delta f$ values of $(5.6\%,4.3,\%3.5\%,1.7\%,-0.4\%,-2.8\%)$; this decreasing trend is in good agreement with an OOD forecast, where the different datasets' distributions share a degree of similarity, which here can be interpreted as the purity value. The closer our distribution is to the maximally mixed states, the lower the reconstruction ability of the protocol, because the model was trained only on the pure states.

%
As long as we use only one inference step, this protocol can obtain improvements in $\delta f$ up to a noise level $p = 0.15$, in the best case. 
In the next section, we improve the current OOD and ID tests by implementing several inference loops.


In this section, we show that it is possible to obtain a further improvement of the OOD reconstruction presented in Fig.~\ref{fig:mixed states} by reusing the network several times in a row, intertwining Eq.~\eqref{den} with Eq.~\eqref{proj}. We can design two different approaches to this, the fixed-point and the simple looping. In this section, we elucidate the fixed-point methodology, which demonstrates superior efficacy. The results pertaining to the simple looping approach are delineated in the Appendix~\ref{app:simple loop}.

\paragraph*{ The fixed-point loop.---} The first approach draws inspiration from the fixed-point theorem. The idea is to reuse the network several times and apply a projection in the feasible set at each application (step). In so doing, our inference process can be described as
\begin{eqnarray}
     \label{eq:fix point inference}
     \vec{c}_{denproj} &=& \Big(\mathcal{P}_\Omega\circ f_{dnn}\Big)^{\circ n_l}[\vec{c}_{cs}].
\end{eqnarray}
\begin{figure}[ht!]
    \centering
    \includegraphics[width = 0.95\linewidth]{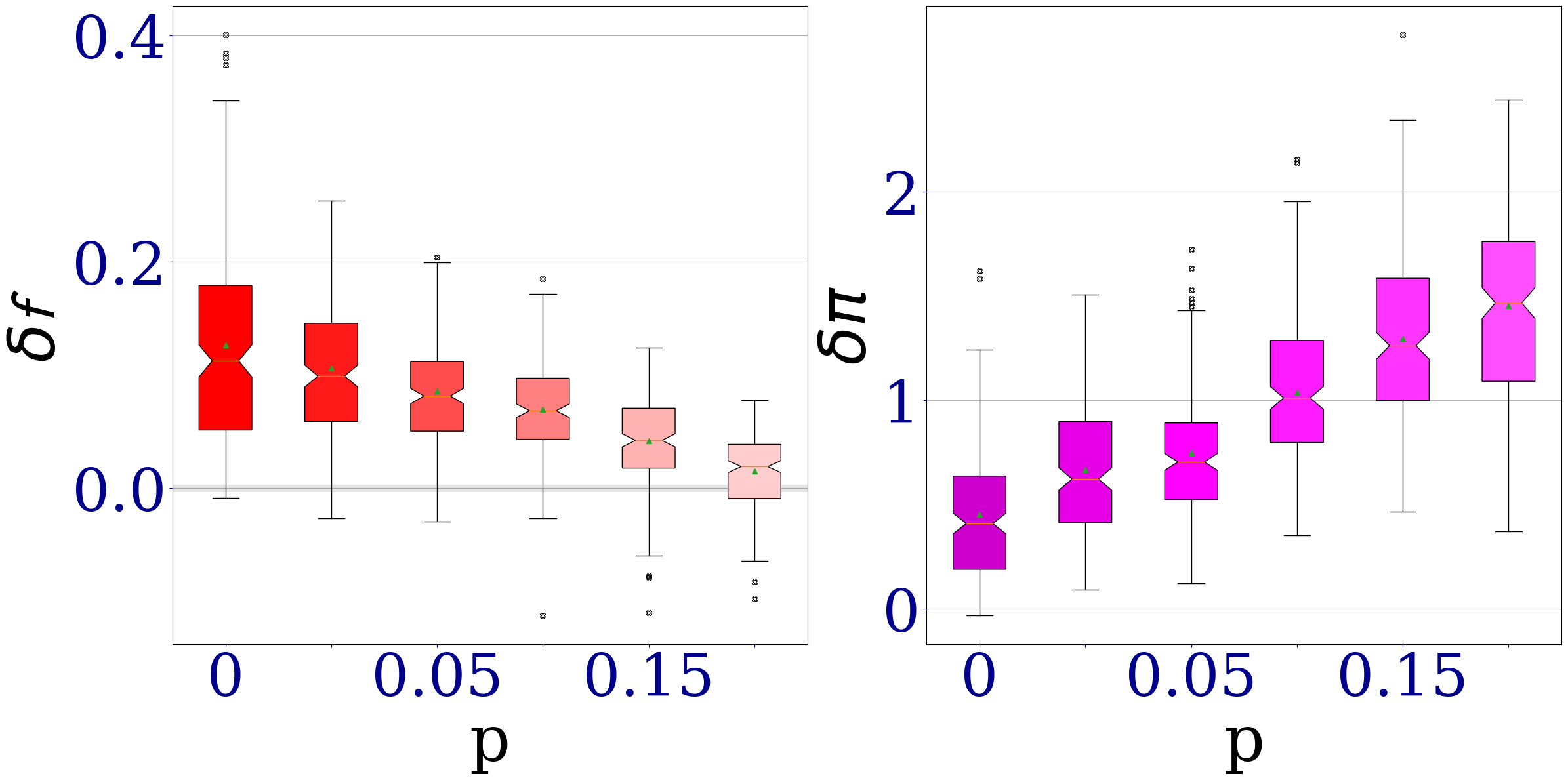}
    \caption{OOD test for the same test dataset used to realize Fig.~\ref{fig:mixed states}, but with the implementation of the fixed-point inference described in Eq.~\eqref{eq:fix point inference}, taking $n_l = 3$. Thanks to this inference strategy, we can achieve a fidelity gain (left) of $\geq 11\%$ for a depolarization strength of $p=0$; an improvement of $\sim 5.5\%$ compared to standard inference with a single inference step provided in the left panel of Fig.\,\ref{fig:mixed states}. Interestingly, the improvement is generalized to all the values of depolarization $p$ considered.}
    \label{fig:fixed point}
\end{figure}

In the left panel of Fig.~\ref{fig:fixed point}, for $n_l = 3$, we can easily notice that this inference strategy can clearly improve $\delta f$ of our outcome reconstructions, obtaining a $11\%$ gain for the ID set, that is, the zero depolarization value. In detail, for the $\delta f$ figure of merit, we obtain an average improvement of $(11\%,10\%,8.2\%,6.7\%,4.1\%,1.4\%)$ each. It should be noted that this approach can still bring gains for a depolarization noise strength of $p=0.2$.

\subsection{Range of applicability}
\begin{figure}[ht!]
    \centering
    \includegraphics[width = 0.85\linewidth]{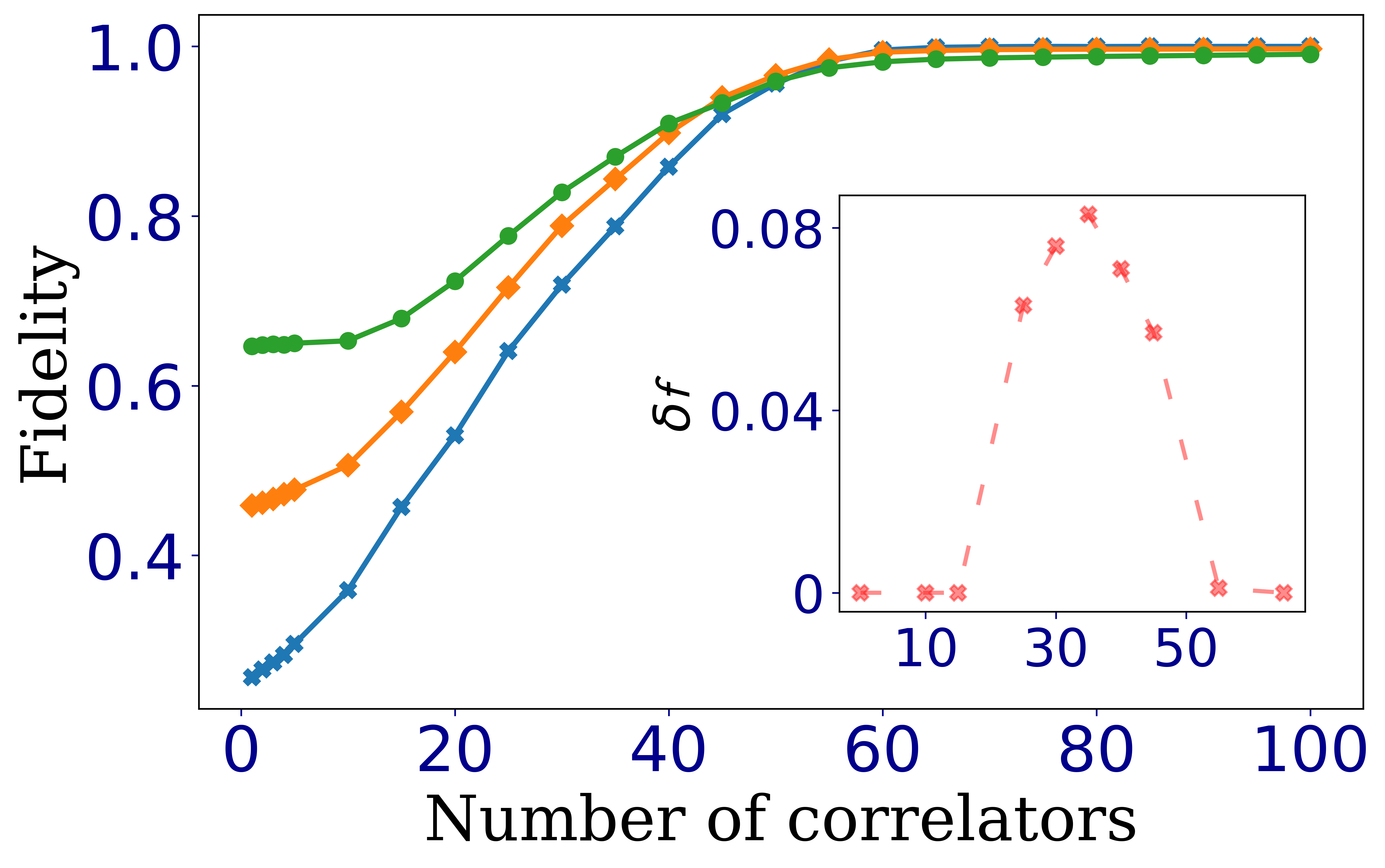}

    \caption{(Main image) mean fidelity as a function of the number of correlators. Each point is the average over $100$ random correlators and $10$ random states. We have the noise $p=0$ (blue), $p=0.05$ (orange), and $p=0.2$ (green). (Inset) $\delta f$ for 1000 Haar non-depolarized states, as a function of the number of correlators. As expected, the protocol gain goes to zero for a low number of correlators, and when it closes the upper bound of 64 as well.
    }
    \label{fig:range of applicability}
\end{figure}

We conclude our analysis by investigating the range of applicability of our method.
%
%
For $K$ sufficiently large, CSR \eqref{cs} alone already gives fidelity $\approx 1$, leaving no room for better options.
On the other hand, when $K$ is too small, the CS input data are informationally too poor to allow one to perform an effective ML training.
This is shown in Fig.~\ref{fig:range of applicability},
where we report the mean fidelity of the CSR as a function of the number of correlators measured. We choose $100$ sets of $N$ random correlators for $N \in [1, 100]$, together with
$\delta f$ as a function of $K$ for pure states, in the inset. 
In the main image, for each set, we reconstruct $10$ random states of the form of Eq.~\eqref{eq:global depo} with $p=0$, $0.05$, $0.2$.
We observe that there is a large difference in the quality
of the reconstruction at the beginning, but for around 40
correlators, the difference stops being significant.

Considering $\delta f$ in the inset, below approximately 20 correlators, the quality of the reconstruction is too poor, making it impossible for the network to make improvements. Vice versa, while approaching the upper bound $K_{cs}\simeq 64$, the quality of the CS reconstruction is already close to one.

\section{Conclusions}\label{sec:conc}
%

Quantum compressed sensing is the fundamental tool for the experimental reconstruction of low-rank density matrices. However, in informationally incomplete scenarios, there are many possible states compatible with the experimental results, and the result given by CS might not be the most accurate. In this work, we introduced a tomographic protocol that concatenates CS reconstruction with a supervised deep neural network architecture followed by a projection in the feasible set. This procedure is able to denoise the CS reconstruction, delivering a quantum state that is, on average, closer to the original experimental state.

The neural network encodes information about the distribution from which the states come, allowing better tomographic reconstruction without requiring additional experimental data. This improvement is evident in our various numerical simulations of Haar random states, which demonstrate an increase in the average fidelity of up to $4.7\%$. Taking advantage of the flexibility offered by deep learning models, we can obtain a two-fold advantage. First, we can design an inference step inspired by the fixed-point theorem, that can further improve the performance of the tomography: by applying the denoising procedure multiple times to the CS reconstruction, we achieve even better results, with an average fidelity gain of $11\%$ for pure states. 
Second, using an out-of-distribution detection argument, we can extend the use of the protocol also to mixed state reconstruction, highlighting the generalization properties of our method. Importantly, 
taking advantage of the OOD paradigm, we can avoid to retrain a supervised network each time for different depolarizing noise, extending our protocol also to mixed quantum state reconstruction without need of any prior knowledge on the target states.

In the tomographic method, we assume a perfect estimation of the correlators, meaning that we are operating in an experimental regime where many shots are used to estimate each correlator. For future developments, we leave the study of the denoising procedure in situations where measurements are imprecise, e.g. standard sampling noise takes place. Lastly, the method could be extended to quantum channel and quantum detector tomography. 
\newline
Data and code are available at \href{https://github.com/AdriQD/OutOfDistribution-CSDNN}{github CS-DNN}


 

\vspace{2cm}
\paragraph*{Acknowledgements.---}
This work was supported by the Government of Spain (Severo Ochoa CEX2019-000910-S, FUNQIP and European Union NextGenerationEU PRTR-C17.I1), Fundació Cellex, Fundació Mir-Puig, Generalitat de Catalunya (CERCA program), the AXA Chair in Quantum Information Science, the ERC AdG CERQUTE and the PNRR MUR Project No. PE0000023-NQSTI.
\\
We also acknowledge support from: 
Europea Research Council AdG NOQIA; 
MCIN/AEI (PGC2018-0910.13039/501100011033,  CEX2019-000910-S/10.13039/501100011033, Plan National FIDEUA PID2019-106901GB-I00, Plan National STAMEENA PID2022-139099NB, I00,project funded by MCIN/AEI/10.13039/501100011033 and by the “European Union NextGenerationEU/PRTR" (PRTR-C17.I1), FPI); QUANTERA MAQS PCI2019-111828-2);  QUANTERA DYNAMITE PCI2022-132919, QuantERA II Programme co-funded by European Union’s Horizon 2020 program under Grant Agreement No 101017733);
Ministry for Digital Transformation and of Civil Service of the Spanish Government through the QUANTUM ENIA project call - Quantum Spain project, and by the European Union through the Recovery, Transformation and Resilience Plan - NextGenerationEU within the framework of the Digital Spain 2026 Agenda;
Fundació Cellex; Fundació Mir-Puig; 
Generalitat de Catalunya (European Social Fund FEDER and CERCA program, AGAUR Grant No. 2021 SGR 01452, QuantumCAT \ U16-011424, co-funded by ERDF Operational Program of Catalonia 2014-2020); 
Barcelona Supercomputing Center MareNostrum (FI-2023-1-0013); 
Funded by the European Union. Views and opinions expressed are, however, those of the author(s) only and do not necessarily reflect those of the European Union, European Commission, European Climate, Infrastructure and Environment Executive Agency (CINEA), or any other granting authority.  Neither the European Union nor any granting authority can be held responsible for them (EU Quantum Flagship PASQuanS2.1, 101113690, EU Horizon 2020 FET-OPEN OPTOlogic, Grant No 899794),  EU Horizon Europe Program (This project has received funding from the European Union’s Horizon Europe research and innovation program under grant agreement No 101080086 NeQSTGrant Agreement 101080086 — NeQST); 
ICFO Internal “QuantumGaudi” project; 
European Union’s Horizon 2020 program under the Marie Sklodowska-Curie grant agreement No 847648;  
“La Caixa” Junior Leaders fellowships, La Caixa” Foundation (ID 100010434): CF/BQ/PR23/11980043.

\medskip

\bibliography{main}

\clearpage
\appendix

\section{Random choice of correlators}
\label{app:pseudorandom}

To demonstrate that the protocol is also suitable for the choice of some correlators other than those provided by the homogeneous settings \eqref{homo-setts}, here we test it using a set of $|\mathcal{I}|=30$ randomly generated correlators, showing that the improvements are not limited to specific choices.

\begin{figure}[ht!]
    \centering
    \includegraphics[width = 0.85\linewidth]{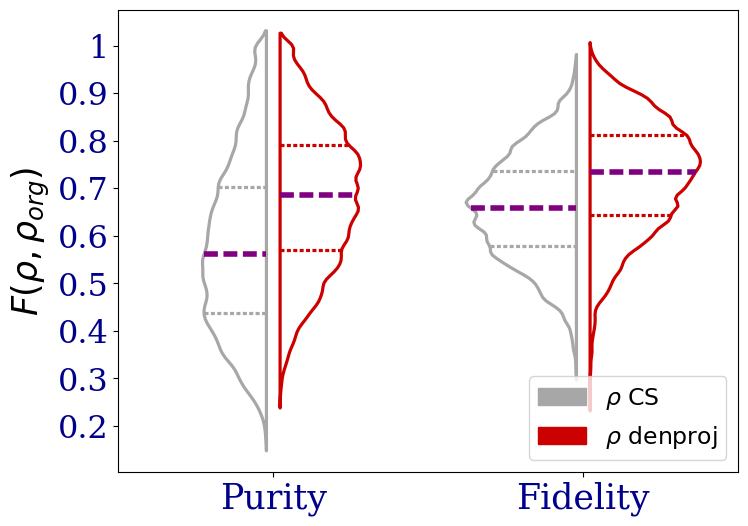}
    \caption{ Distributions comparison for the purity $\pi$ and quantum Fidelity $F$ for 6000 Haar test states. Horizontal dotted purple lines show the average value, dotted gray ones the distributions quartile. Throughout the protocol we employ a set $\mathcal{I}$ of 30 random correlators. With this set, the DL model can offer an average fidelity improvement of $\geq 6\%$ and $\sim 13\%$ in purity. }
    \label{fig:random stats}
\end{figure}

As we can see in Fig.~\ref{fig:random stats}, using the random set, we can achieve a fidelity gain of $\sim 6\%$, although the average fidelity reconstruction obtained over the test set is lower. This is expected, with the total number of correlators being lower compared to the homogeneous set.
This test shows that the approach is suitable for a generic choice of measurements.

\section{Simple loop inference}
\label{app:simple loop}
A quick alternative for the inference looped phase is a simpler approach, where we skip the projection in the feasible set until the very end of the inference, formally
\begin{equation}
\label{eq:simple looping inference}
    \vec{c}_{denproj} =  \mathcal{P}_\Omega\circ {(f_{dnn})}^{\circ n_l}[\vec{c}_{cs}]\,.
\end{equation}

\begin{figure}[h!] 
    \centering
    \includegraphics[width = 0.95\linewidth]{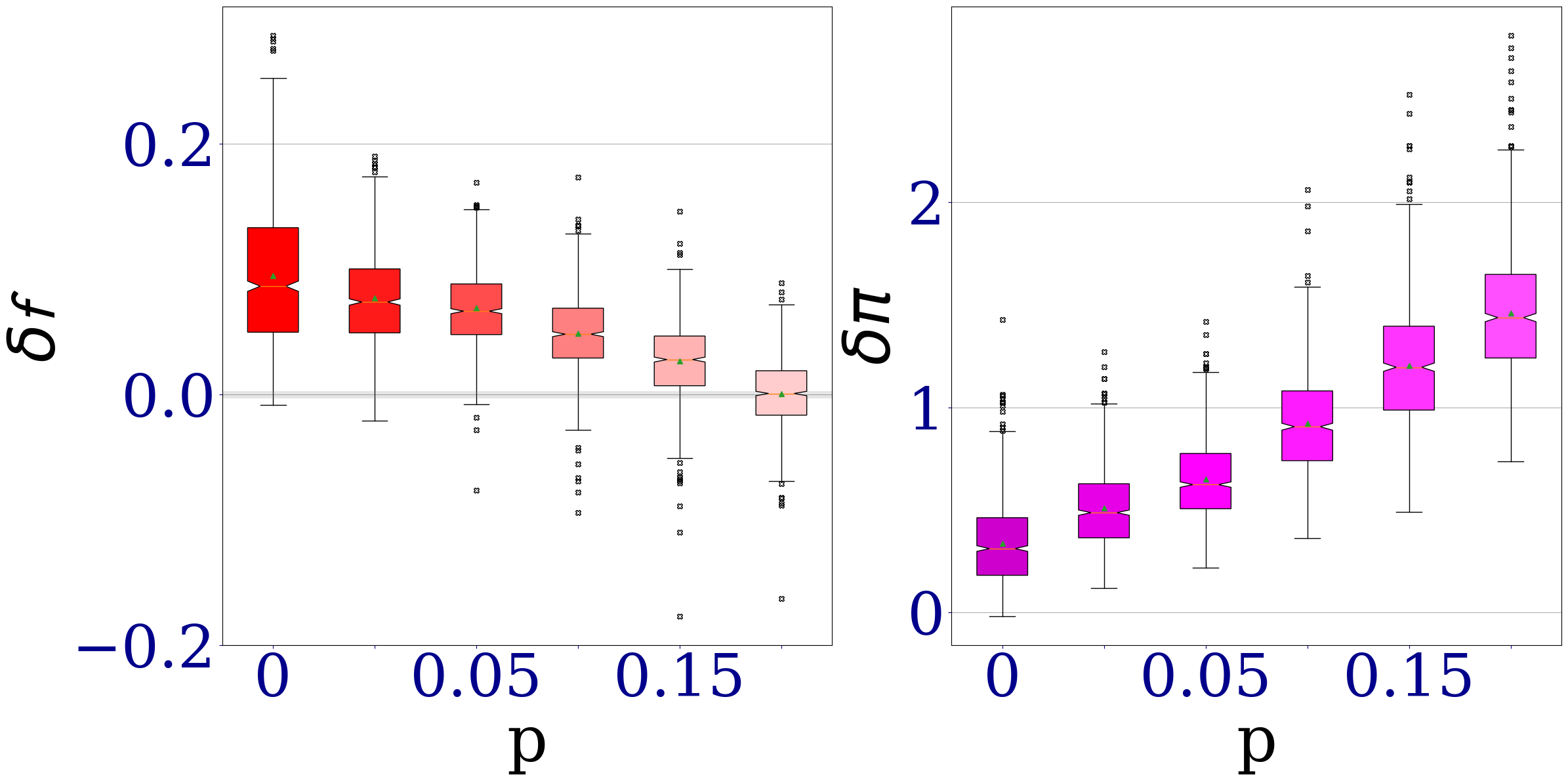}
    \caption{ OOD test for the same test dataset used to realize Fig.~\ref{fig:mixed states}, but with implementing the simple loop inference described in Eq.~\eqref{eq:simple looping inference}, where the projections step appears only in the last stage.
    With this simpler approach, we can obtained a general improvement in the range $\sim 3-5\%$ for $\delta f$ compared to Fig.~\ref{fig:mixed states}, e.g. obtaining an average value of $\delta f \sim 8.4\%$ for non-depolarized states. Overall, compared to Fig.~\ref{fig:mixed states} we can still obtain an improvement, but of lower quality compared to the fixed-point type inference in Fig.~\ref{fig:fixed point}.}
    \label{fig: looped inference}
\end{figure}

This inference approach allows us to figure out the real ability of this specific specific architecture alone to reconstruct our incomplete data because different models and strategies can lead to different results. We found that three recursive applications suffice for our data, while for a higher number of loops, we witness a degradation of the reconstruction quality; see the Appendix~\ref{app:stopping rule} for details.

With this approach, taking $n_l=3$, we obtain an average $\delta f$ of $(8.4\%,7.0\%,6.4\%,4.5\%,2.5\%,0\%)$.
Although the overall quality is lower than that of the fixed point, this simple-loop inference also brings about a general improvement in the outcomes.

\section{The iterations stopping rule}
\label{app:stopping rule}

\begin{figure}[h!]
    \centering
    \includegraphics[width = 0.9\linewidth]{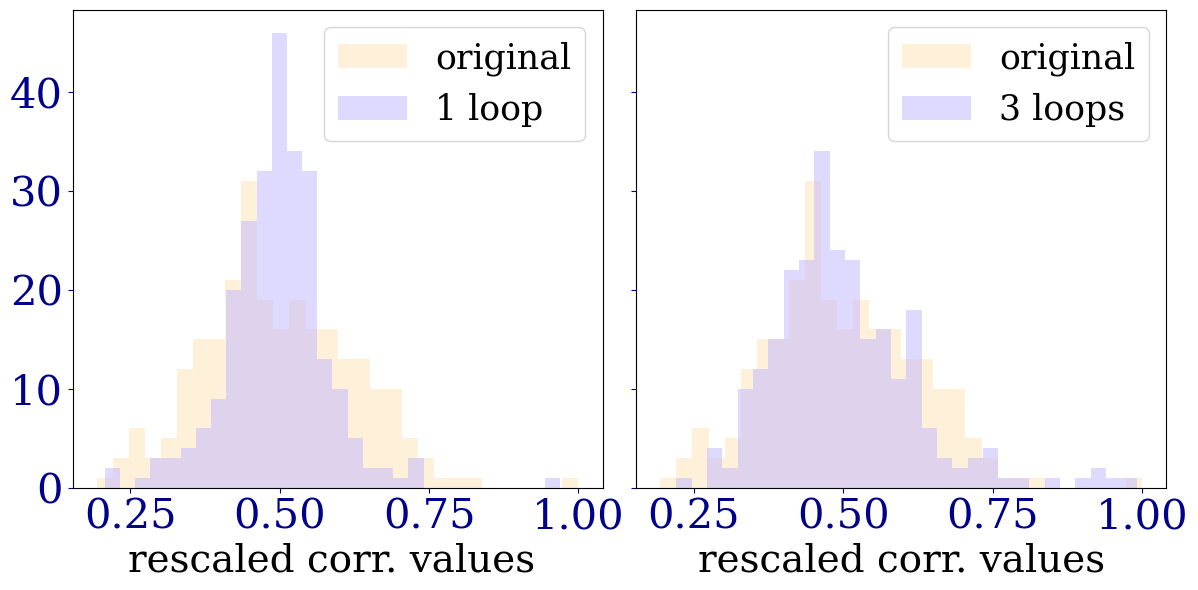}
    \caption{Example of a single input data distribution for the original and (left) for one single step of inference and (right) for a total of three inference steps with the simple-loop, without the final projection in the feasible set. We can see that the looped-inference action ensures that the distribution standard deviation grows, obtaining a better overlap with the original one, thus, improves the outcomes. }
    \label{fig:loops}
\end{figure}

A big advantage of our looped inference approach is that it allows us to optimize the network with synthetic data \textit{beforehand}, in a completely agnostic way to the physical source of noise, at least for the depolarization channel.
In practice, it is possible to figure out a sufficient number of loops by performing a numerical analysis on synthetic data and making use of it.
Formally, we can retrieve this relationship from the MSE cost function by rewriting $MSE = \mathbf{E}(\hat\theta_m -\theta)^2= {\rm Bias}(\theta_m) + {\rm Var}(\theta_m)$ (see \cite{Goodfellow-et-al-2016}, pag.127), whose optimization is dictated by the \textit{ bias-variance trade-off} principle. This principle states that when the bias is lower, the variance has to increase and vice versa. 
In our looped inference, we can reduce the bias error; therefore, we pay attention to variance behaviors to offer a general rule of thumb.
Detailly:

\begin{enumerate}
    \item we apply the network to our test set for different number of loops $l$, generating a collection of rescaled correlators $\vec{c}_{den}$
    \item we evaluate the averaged standard deviation $\sigma^l_i :=\sigma^l(\vec{c}_{nn,i})$  of the entire test set, at each loop $l$ 
    \item we measure the gain at each step $\Delta_{l}^{l+1} = \frac{\mathbb{E}(\sigma^{l+1}_i)-\mathbb{E}(\sigma^l_i)}{\mathbb{E}(\sigma^l_i)}$ 
\end{enumerate}

The standard deviation values behave pseudo logarithmically, according to our interpretation in Fig.~\ref{fig:schema}. In numerical analysis, a useful stopping point can be found when the standard deviation gain between two adjacent steps reaches $\sim 5-6\%$. Below this threshold, increasing the number of loops may reduce $\delta f$.



\section{Algorithms details}

\paragraph*{ Network architecture.---} As mentioned, our architecture draws inspiration from a subclass of models that are known by different names. Its realization combines the best of two operations of opposite ability, convolutions, suitable for identifying \textit{ local} features within the data \cite{CNN-Peng2021}, due to the point-wise structure of the convolution function and an encoder transformer, efficient in pinpointing \cite{TRANSFORMER-Guo2022} characteristics from the data and \textbf{superior} to convolutional neural network on generalization on out-of-distribution sample reconstruction \cite{TRANSFORMERvsCNN}. 
In this architecture, the CNN input layer behaves like a first encoder block, which realizes several representations of the same input. These new representations are elaborated by a second encoder, the encoder-transformer, that uses this information to adjust the numerical errors from the CS reconstructions. 
Lastly, a final CNN layer realizes a decoder layer that outputs our reconstructions $\vec{c}_{den}$. 

\paragraph*{ Hardware \& software specifications.---} All numerical experiments were carried out using a Linux-based system, with 128 CPU and an NVIDIA A100 80GB GPU, Pythorch version 2.1 and nvidia CUDA 12.1. For data generation, convex optimization software with MOSEK optimizer was used.

\end{document}